\documentclass[a4paper, conference]{IEEEtran}

\usepackage{graphicx}
\usepackage{epsfig}
\usepackage[ansinew]{inputenc}
\usepackage{amsmath}
\DeclareMathOperator*{\argmin}{argmin} 
\DeclareMathOperator*{\mean}{mean} 
\usepackage{amssymb}
\usepackage{subfig}
\usepackage{epstopdf}
\usepackage{bm}
\usepackage{dblfloatfix}
\usepackage{tabulary}
\usepackage{multirow}
\usepackage{flushend}
\usepackage{booktabs}

\usepackage[shortcuts,acronym, nomain]{glossaries}

\usepackage{floatrow}
\makeglossaries 

\newacronym{urllc}{URLLC}{Ultra Reliable Low Latency Communication}
\newacronym{ml}{ML}{Machine Learning}
\newacronym{physec}{PHYSEC}{Physical Layer Security}
\newacronym{sdr}{SDR}{Software Defined Radio}
\newacronym{mbb}{MBB}{Mobile Broadband}
\newacronym{mac}{MAC}{Message Authentication Code}
\newacronym{mcmtc}{MC-MTC}{Mission Critical Machine Type Communication}
\newacronym{ofdm}{OFDM}{Orthogonal Frequency Division Multiplexing}
\newacronym{rrm}{RRM}{Radio Resource Management}
\newacronym{ai}{AI}{Attack Intensity}
\newacronym{roc}{ROC}{Receiver Operating Characteristic}
\newacronym{sgd}{SGDC}{Stochastic Gradient Descent Classifier}
\newacronym{pa}{PAC}{Passive Agressive Classifier}
\newacronym{gnb}{GNBC}{Gaussian Naive Bayes Classifier}
\newacronym{rf}{RFC}{Random Forest Classifier}
\newacronym{kn}{K-NC}{K-Neighbors Classifier}
\newacronym{svc}{SVC}{Support Vector Classifier}
\newacronym{lda}{LDAC}{Linear Discriminant Analysis Classifier}
\newacronym{aes}{AES}{Advanced Encryption Standard}
\newacronym{cmac}{CMAC}{cipher based MAC}
\newacronym{hmac}{HMAC}{hash based MAC}
\newacronym{rssi}{RSSI}{Received Signal Strength Indicator}
\newacronym{mimo}{MIMO}{Multiple Input Multiple Output}
\newacronym{siso}{SISO}{Single Input Single Output}
\newacronym{tcp}{TCP}{Transmission Control Protocol}
\newacronym{usrp}{USRP}{Universal Software Radio Peripheral}
\newacronym{fft}{FFT}{Fast Fourier Transform}
\newacronym{los}{LOS}{Line of Sight}
\newacronym{simo}{SIMO}{Single Input Multiple Output}
\newacronym{cfo}{CFO}{Carrier Frequency Offset}
\newacronym{phy}{PHY}{Physical Layer}

\begin{document}

\title{Supervised Learning for Physical Layer based Message Authentication in URLLC scenarios}

\author{\IEEEauthorblockN{Andreas Weinand, Raja Sattiraju, Michael Karrenbauer, Hans D. Schotten}
\IEEEauthorblockA{Institute for Wireless Communication and Navigation\\
University of Kaiserslautern, Germany\\
Email: \{weinand, sattiraju, karrenbauer, schotten\}@eit.uni-kl.de
}
}

\maketitle


\begin{abstract}
PHYSEC based message authentication	can, as an alternative to conventional security schemes, be applied within \gls{urllc} scenarios in order to meet the requirement of secure user data transmissions in the sense of authenticity and integrity. In this work, we investigate the performance of supervised learning classifiers for discriminating legitimate transmitters from illegimate ones in such scenarios. We further present our methodology of data collection using \gls{sdr} platforms and the data processing pipeline including e.g. necessary preprocessing steps. Finally, the performance of the considered supervised learning schemes under different side conditions is presented.
\end{abstract}

\section{Introduction}
\label{intro}
{\let\thefootnote\relax\footnote{This is a preprint, the full paper will be published in Proceedings of IEEE 90th Vehicular Technology Conference: VTC2019-Fall, \copyright 2019 IEEE. Personal use of this material is permitted. However, permission to use this material for any other purposes must be obtained from the IEEE by sending a request to pubs-permissions@ieee.org.}}
The upcoming demand for automation in our everydays life is increasing continuously. Many \gls{urllc} services, such as automated traffic systems, telesurgery and many more are not far from becoming a reality or already are. In order to meet the high requirements for this type of services, new approaches, especially whithin the field of information security, need to be considered. Beside authenticity, integrity and confidentiality of the data to be transmitted over wireless links have to be ensured in order to prohibit a wide range of passive and active cyber attacks. Confidentiality can be ensured by encryption of messages, e.g. using an \gls{aes} \cite{AES_standard} cipher suite. This operation does not produce any overhead in message size (except zero padding). Whereas in case of message authenticity and integrity schemes some amount of overhead is generated in the sense of authentication tags, e.g. when a \gls{mac} is used or cryptographic certificates, which both need to be attached to a message respectively.
E.g., the recommendation of the Internet Engineering Task Force is to either use a \gls{cmac}, e.g. based on a \gls{aes}-128 block cipher, or a \gls{hmac} which is based on a cryptographic hash function. For \gls{aes}-128 based \gls{cmac} a maximum shortening to 64 Bit is recommended \cite{RFC_CMAC.2006}, while for \gls{hmac} a minimum \gls{mac} size of 80 Bit is recommended \cite{RFC_HMAC.1997}. If we now assume that the payload of a \gls{urllc} message in the uplink has a length of 400 Bit (the dimension of this assumption is e.g. confirmed by \cite{Bockelmann2017} and \cite{Osman.2015}), then the overhead and with this the additional latency introduced in case of a \gls{aes}-128 based \gls{cmac} of $64$ Bit length would be $13.8\%$. Another important issue is, that key based schemes such as \gls{mac} are only able to protect the message payload from the mentioned attacks. An attacker is still able to perform attacks such as address spoofing, or even worse, record a message and replay it after a while (if counters are used to prohibit that, additional overhead is added as well). Due to these drawbacks, \gls{physec} based message authentication can be used alternatively. Hereby, the physical properties of the radio link signal are taken into account in order to identify, by which user a message was originally transmitted and whether it was modified during the wireless transmission. As we focus on \gls{urllc} here, it can be assumed that frequent and periodic data transmissions (e.g. within the order of $10$s of ms \cite{Bockelmann2017}, \cite{Osman.2015}) and consequently channel estimation at the same rate is available. For experimental evaluation, we consider an \gls{ofdm} system and based on the respective frequency domain channel estimations, we decide from which source a received data packet was transmitted.\\
The remainder of the work is organized as follows. In section \ref{related work} we give a short overview on related work with respect to previous considered approaches and in section \ref{system_model_section} we describe the system model. The approach of \gls{phy} based message authentication is presented in section \ref{physec_section} and the respective application of \gls{ml} in section \ref{ml_section}. In section \ref{results} we present the results of our work in form of an experimental evaluation of the considered approach and section \ref{CONC} finally concludes the paper. 

\section{Related Work}
\label{related work}
Several approaches on exploiting characteristics of the \gls{phy} in order to derive security functionalities (\gls{physec}) have been investigated recently. In \cite{Jorswieck.2015} an overview is given on \gls{physec} functions derived from the wireless channel. While many works have focused on extracting secret keys (e.g. to be used for encryption and decryption of user data), such as \cite{Guillaume.}, \cite{Zenger.2014}, \cite{Ambekar.2012b}, the focus of our work is to identify the original transmitter of data packets by exploiting the related \gls{phy} metadata. 

First, statistical methods were used for \gls{physec} based authentication, e.g. \cite{Xiao.2007}, where an approach based on hypothesis testing is presented for static scenarios. It is later extended to mobile and time-variant scenarios in \cite{Xiao.2008}. Though the results of these works are plausible, they are obtained from simulations of the wireless channel only and therefore, it could not be proven, whether the methods also work for real world channels. Our work is in contrast considering measurements from real world wireless channels. Additionally, such testing methods require proper setting of the detection thresholds. In order to overcome this, \gls{ml} based methods were proposed in recent years for \gls{physec} based authentication as an alternative. In \cite{Pei.2014}, two approaches based on Support Vector Machines and Linear Fisher Discriminant Analysis respectively, are presented that yielded acceptable results. A Gaussian Mixture Model based technique in combination with exploitation of the channel responses for different antenna modes is considered in \cite{Gulati.2013}. Due to these promising results, the goal of our work is to further investigate \gls{ml} based identification methods, especially such using supervised learning.

In previous works, various \gls{phy} features ranging from hardware characteristics such as \gls{cfo} to channel characteristics such as impulse or frequency responses have been used in order to identify the associated transmitters of data packets. E.g. the approach in \cite{Tugnait.2010} is based on time domain channel estimation within a single carrier system. In contrast to that, we will focus on multi carrier systems in our work as they are more common nowadays. The authors in \cite{Caparra2016} consider a cellular Internet-of-Things system and propose a scheme for channel based message authentication of nodes with the help of anchor nodes, which are assumed to have set up trust to the concentrator node before using \gls{rssi} values. Also, in \cite{Shi.2013} an \gls{rssi} based approach for body area networks is presented. As we only consider single links, a single and onedimensional feature might not be sufficient in order to reliable authenticate transmitters of the corresponding messages. Therefore, we will use the channel frequency response within \gls{ofdm} systems in order to identify respective transmitters. The work in \cite{Refaey.2014} considers a multilayer approach based on \gls{ofdm} to guarantee authentication of \gls{tcp} packets. In \cite{Baracca2012}, a \gls{mimo} \gls{ofdm} system is considered and a generalized likelihood ratio test is used as detection method. Further, the authors derive information theoretic bounds for the channel based detection problem. In \cite{Forssell2019a} and \cite{Forssell2019b} the authors use a \gls{simo} system in order to detect the transmitter of data packets based on the respective \gls{los} components. They further develop models in order to estimate the delays within message delivery in case of false alarms or misdetection and considere a variety of attack strategies, such as disassociation or sybil attacks. 

\section{System Model}
\label{system_model_section}
In this section, the system concept, as well as the channel estimation technique, which is used to derive the \gls{phy} metadata in our work, are introduced. Further, the attacker model is presented and we derive possible attack scenarios with respect to \gls{urllc}.

\begin{figure}[h]
\centering
\includegraphics[width=0.9\textwidth]{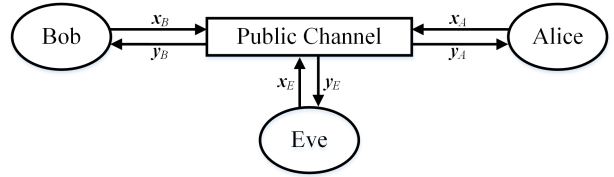}
\caption{System Model}
\label{system_model}
\end{figure}

Here, we consider a wireless (public) channel between two users, Alice and Bob, who want to exchange authenticated messages with each other over this channel. A third user Eve, who is an adversarial user, tries to inject illegal messages masqueraded as one of the legal users Alice or Bob. This can e.g. be done by reusing the radio ressources allocated to either Alice or Bob (in case of centralized \gls{rrm}) or by spoofing their addresses (in case of competetive/decentralized \gls{rrm}). Fig. \ref{system_model} shows the possible message flows over the wireless channel for all considered users. Messages transmitted by users are denoted with $\boldsymbol{x}_u$ and messages received by users with $\boldsymbol{y}_u$, where $u$ denotes user Alice ($u=A$), Bob ($u=B$) or Eve ($u=E$).

\subsection{Physical Layer Metadata}
In \gls{physec} schemes, \gls{phy} protocol metadata is used in order to provide lightweight security measures. This metadata can either originate from the transceiver hardware (e.g. radio fingerprints or clock skew) or can originate from characteristics of the wireless channel (e.g. \gls{rssi} or channel estimation data). In the first case, it is already challenging for an attacker to spoof these characteristics by using default hardware (off-the-shelf). On the other hand, this also requires an additional effort for the legitimate receiver in order to detect these characteristics and increases the receiver complexity and cost. For the second case, the respective data is typically available and accessible in any radio protocol. Therefore, in our work, we focus on channel based metadata and especially channel estimation data, which is e.g. computed at the receiver side in form of a frequency domain channel estimation (\gls{ofdm} based systems). The received signal at a user $u_i$ in such systems is
\begin{equation}
\boldsymbol{y}_{u_i}=\boldsymbol{H}_{u_iu_j}{\boldsymbol{x}_{u_j}}+{\boldsymbol{n}_{u_iu_j}}
\label{eq:}
\end{equation}
with
\begin{equation}
\boldsymbol{H}_{u_iu_j}=[{h}^0_{u_iu_j},\ldots, {h}^{M-1}_{u_iu_j}]
\label{eq:}
\end{equation}
being the $M$-dimensional channel matrix of the frequency-selective \gls{siso} fading channel between user $u_i$ and $u_j$ (transmitter) and ${h}^\ell_{u_iu_j}$ being the complex gain of the $\ell$-th sample in frequency domain ($\ell=0,\ldots,M-1$). It can be estimated as 
\begin{equation}
\boldsymbol{\hat{H}}_{u_iu_j}=\boldsymbol{H}_{u_iu_j}+\boldsymbol{\epsilon}_{u_iu_j}
\label{eq:}
\end{equation}
by user $u_i$. Due to noise $\boldsymbol{n}_{u_iu_j}$ which is modelled as a gaussian random variable with zero mean and variance $\boldsymbol{\sigma^2}_{\boldsymbol{n}_{u_iu_j}}$ the channel estimation is not perfect and errors $\boldsymbol{\epsilon}_{u_iu_j}$ occur.

\subsection{Attacker Model}
\label{attacker_model}
A typical scenario for an attacker Eve is, that he is at a spatially different location compared to Bob and Alice (we assume a distance of more than the wavelength of the transmitted signal respectively) and uses advanced equipment such as directed antennas and high sensitivity receivers in order to maximize the coverage to his benefits. We also assume perfect knowledge of the underlying communication protocol at Eve to run active attacks such as masquerade attacks, replay attacks or address spoofing attacks by introducing messages $\boldsymbol{x}_{\mbox{\scriptsize E}}$.
It is not assumed that Eve is gaining physical access to Alice or Bob to accomplish invasive attacks such as hardware modification. Further, other active attacks such as Denial-of-Service attacks due to jamming are not considered as well. It is assumed, that the legal communicating participants Bob and Alice have already carried out initial user authentication to each other and have set up trust in a secure way by using cryptographic certificates or \gls{mac}s. Attacks on the initial authentication stage are not considered. Further, we assume that Eve is, depending on the current attack strategy, spoofing on transmission slots which are actually allocated to Bob by using a higher transmit power in order to make Alice overhear Bobs transmissions. In order to successfully carry out attacks, the receive power of Eves transmissions has actually to be higher, compared to the receive power of Bobs transmissions at Alice.

\section{\gls{physec} based message authentication}
\label{physec_section}
One of the main characteristics, which can be exploited in \gls{physec} based schemes, is the spatial decorrelation property of the wireless channel. It denotes, that channel characteristics are varying based on the location of a transceiver. This holds already true for small variations in the spatial domain which are in the order of the wavelength of the transmitted signal, due to positive and negative superposition of wireless signals, caused by e.g. reflection and scattering.  
In our case, let us assume, that Bob wants to transmit authenticated messages to Alice. Therfore, within a first step, Alice will estimate the channel as $\boldsymbol{\hat{H}}_{\mbox{\scriptsize AB}}{(k)}$ when receiving message $\boldsymbol{y}_{\mbox{\scriptsize A}}{(k)}$ at times $k=0,...,T-1$ from user Bob (transmits $\boldsymbol{x}_{\mbox{\scriptsize B}}{(k)}$). In order to prohibit any spoofing attacks by Eve during time $k=0,...,T-1$, the messages $\boldsymbol{x}_{\mbox{\scriptsize B}}{(k)}$ sent by Bob at times $k=0,...,T-1$ need to contain cryptographic information from higher layers (e.g. certificates or authentication tags). Next, if Alice wants to authenticate Bob as the transmitter of further received messages $\boldsymbol{y}_{\mbox{\scriptsize A}}{(k)}$ at time $k \ge T$, this can be achieved by comparing the new channel estimations $\boldsymbol{\hat{H}}{(k)}, k \ge T$ to the previous collected channel estimations $\boldsymbol{\hat{H}}_{\mbox{\scriptsize AB}}{(k)}, k=0,...,T-1$. Based on these assumptions, there are two hypothesis about the origin of the respective estimated channel $\boldsymbol{\hat{H}}{(k)}, k \ge T$ and consequently the respective payload of it, either
\begin{equation} \label{eq1}
\begin{split}
\mathcal{H}_0&: \boldsymbol{\hat{H}}{(k)} \ \mathrm{due\ to\ } \boldsymbol{x}_{\mbox{\scriptsize B}}{(k)},\ \mathrm{or\ } \\
\mathcal{H}_1&: \boldsymbol{\hat{H}}{(k)} \ \mathrm{due\ to\ } \boldsymbol{x}_{\mbox{\scriptsize E}}{(k)}.\
\end{split}
\end{equation}
In case of hypothesis $\mathcal{H}_0$ 
\begin{equation}
\boldsymbol{\hat{H}}{(k)}=\boldsymbol{H}_{\mbox{\scriptsize AB}}{(k)}+\boldsymbol{\epsilon}_{\mbox{\scriptsize AB}}{(k)},
\label{eq:}
\end{equation}  
and in case of hypothesis $\mathcal{H}_1$ 
\begin{equation}
\boldsymbol{\hat{H}}{(k)}=\boldsymbol{H}_{\mbox{\scriptsize EB}}{(k)}+\boldsymbol{\epsilon}_{\mbox{\scriptsize EB}}{(k)}.
\label{eq:}
\end{equation}
Due to temporal variations in wireless channels, e.g. due to doppler fading, the reference channel estimation used for comparison needs to be updated accordingly. Otherwise it is not possible to distinguish between channel variations and messages introduced by an attacker after some point. Therefore, the update interval should be below the channel coherence time $T_c$ of the $\boldsymbol{H}_{\mbox{\scriptsize AB}}$ channel in order to catch up with these variations. Whereas in both cases, the temporal difference between two subsequent channel estimations should always be below $T_c$. 


\section{Supervised Learning}
\label{ml_section}
This section introduces the concept of supervised \gls{ml} based classification. Further, we introduce the necessary data processing steps (pipeline) including data prepocessing, the hyperparameter optimization strategy we applied and how the performance of classifiers for the \gls{physec} based message authentication task can be evaluated.

\subsection{\gls{ml} based Classification}
The \gls{ml} classifiers that were applied to the problem of \gls{physec} based authentication are introduced in detail in section \ref{results} including their hyperparameter spaces. In general in supervised learning based classification, the goal is to learn the function $f(x)=w^Tx+b$ (or its parameters), which maps the input data $x=[x_1,...,n_N], x_k \in \mathbb{R}^d$ to the outputs $x=[y_1,...,y_N], y_k \in \lbrace -1,1\rbrace$ (binary classification), the regularized training error
\begin{equation}
E(w,b)=\frac{1}{N}\sum_{k=1}^{N}L(y_k, f(x_k))+\alpha R(w),
\label{supervised_general}
\end{equation}
has to be minimized. Here, $L$ is the loss function, $R$ is the regularization or penalty term and $\alpha \in \mathbb{R}_0^+$ is a hyperparameter.
  
%
%
%
%

\subsection{Data Processing Pipeline}
Within this subsection, the processing pipeline including functionalities such as data pre-processing, hyperparameter optimization and finally testing the different \gls{ml} models in order to evaluate their performance.

\subsubsection{Preprocessing}
The following preprocessing steps are applied to the validation and testing datasets respectively, in order to fit the classifiers. First, the magnitudes
\begin{equation}
\boldsymbol{{F}}(k) = \big|\boldsymbol{\hat{H}}{(k)}\big|, \ k=0,\ldots,N
\end{equation}
of the complex channel gains are calculated for each data sample $\boldsymbol{\hat{H}}{(k)}$. As it is in some cases necessary to reduce the dimensionality $M$ of the the input feature in order to avoid overfitting or underfitting, we also add a preprocessing step for that purpose. This reduces the number of features of the original sample from $M=48$ to $M_{red} \in [1, 2, 3, 4, 6, 8, 12, 16, 24, 48]$ by either sampling the original feature vector at the respective rate yielding
\begin{equation}
\boldsymbol{{F}}_{red}(k,\ell^*) = \boldsymbol{{F}}(k,\frac{\ell^* M}{M_{red}}), \ \ell^*=0,\ldots,M_{red}-1,
\end{equation}
or by applying a mean function according to
\begin{equation}
\label{mean_reduction}
\boldsymbol{{F}}_{red}(k,\ell^*) = \frac{M_{red}}{M}\sum_{j=\frac{\ell^*M}{M_{red}}}^{\frac{(\ell^*+1)M}{M_{red}}-1}\boldsymbol{{F}}(k,j), \ \ell^*=0,\ldots,M_{red}-1,
\end{equation}
where $\ell^*$ denotes the index of the features in the new feature dimension $M_{red}$ and $\ell$ the index in the original feature dimension.

The next steps involves a conversion of the time series data $\boldsymbol{{F}}_{red}(k,\ell^*)$ into the a format that can be processed by a supervised classifier according to equation \ref{supervised_general}. A sample $\boldsymbol{{F}}_{window}(k)$ at time $k$ now consists of the original (reduced) sample $\boldsymbol{{F}}_{red}(k)$, as well as the $W$ previous samples, yielding
\begin{equation}
\label{window_function}
\boldsymbol{{F}}_{window}(k) = [\boldsymbol{{F}}_{red}(k), \boldsymbol{{F}}_{red}(k-1), \ldots, \boldsymbol{{F}}_{red}(k-W)].
\end{equation}
Next, the data sets are split into training and validation or testing sets, which are used to fit the classifiers (training) and validate (grid search) or test them (performance evaluation). The training set is given by 
\begin{equation}
\boldsymbol{{F}}_{train} = [\boldsymbol{{F}}_{window}(0),\ldots,\boldsymbol{{F}}_{window}(N_{train}-1)],
\end{equation}
and the remaining samples are used for validating or testing respectively. Finally, the training and validation/testing data is normalized. We calculate the mean and variance of the training set, then the training set is transformed to 
\begin{equation}
\boldsymbol{{F}}^{norm}_{train}=\frac{\boldsymbol{{F}}_{train}-mean(\boldsymbol{{F}}_{train})}{std(\boldsymbol{{F}}_{train})},
\label{eq:}
\end{equation}
whereas the validation set (and testing set respectively) is transformed according to
\begin{equation}
\boldsymbol{{F}}^{norm}_{valid}=\frac{\boldsymbol{{F}}_{valid}-mean(\boldsymbol{{F}}_{train})}{std(\boldsymbol{{F}}_{train})}.
\label{eq:}
\end{equation}

\subsubsection{Grid Search}
\label{gridsearch}
In order to find the optimal hyperparameters 
\begin{equation}
\lambda^* = \argmin_{\lambda \in \Lambda} \mean_{x_k^{(valid)}} L(y_k^{(train)}, f_{\lambda}(x_k^{(train)})
\end{equation}
to the given classification task, we use an exhaustive grid search. For each model, possible options for each hyperparameter are given as input and the respective models are fitted and validated on all combination of hyperparameters. For this, we only use a part of the overall $L_{total}=L_{valid}+L_{test}$ data sets, which we call the validation set. It consists of $L_{valid}$ data sets. We use the accuracy metric in order to select the optimal hyperparameter set for each classifier, which is given by
\begin{equation}
\mathcal{P}_A = \frac{1}{N}\sum_{k=N_{train}}^{N-1}y_{pred}(k)-y_{true}(k),
\end{equation}
where $\boldsymbol{y}_{pred}$ is the label vector predicted by the respective classifier (user Bob or Eve is the transmitter) and $\boldsymbol{y}_{true}$ the true label vector.

\subsubsection{Classifier Performance Testing}
After deriving the optimal hyperparameters for each classifier, the remaining $L_{test}$ data sets are used in order to evaluate their performance on the given classification task. For that purpose, beside the accuracy metric, e.g. other metrics such as \gls{roc} curves might be used. 

\section{Performance Evaluation}
\label{results}
In this section, we first describe the process of acquisition of channel estimation data and how the data is further processed in order to evaluate the introduced \gls{ml} classifiers. We present the respective results and discuss them.

\subsection{Data Acquisition}
\begin{figure}[b]
	\centering
	\includegraphics[width=2in]{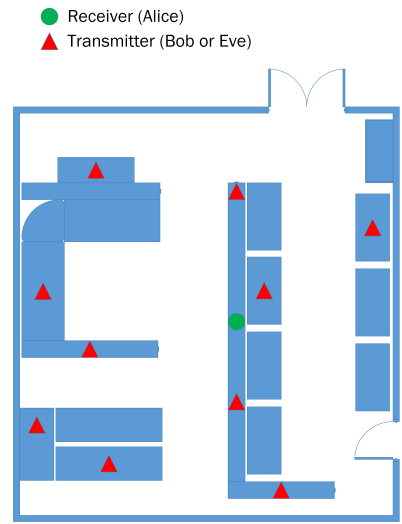}
	\caption{Environment with different Bob and Eve positions}
	\label{environment}
\end{figure}

\begin{table*}[t!]
	\caption{Parameter space of all classifiers candidates}
	\label{param_space}
	\footnotesize
	\centering
	\begin{tabulary}{1.0\textwidth}{|c|c|c|}
		\hline
		Model & Parameter & Options \\
		\hline
		\multirow{7}*{SGDClassifier} & loss & [hinge, log, modified\underline{ }huber, squared\underline{ }hinge, perceptron] \\ 
		& Penalty  & [none, l2, l1, elasticnet] \\
		& alpha & [$10^{-6}$, $10^{-5}$, $10^{-4}$, $10^{-3}$, $10^{-2}$] \\
		& l1\underline{ }ratio & [0, 0.25, 0.5, 0.75, 1]\\
		& max\underline{ }iter & [1, 10, 100, 1000, 10000]\\
		& tol & [$10^{-5}$, $10^{-4}$, $10^{-3}$, $10^{-2}$, $10^{-1}$]\\
		& learning\underline{ }rate & [constant, optimal, invscaling, adaptive]\\
		& eta0 & [0.25, 0.5, 0.75, 1]\\
		\hline
		\multirow{4}*{Perceptron} & penalty & [none, l2, l1, elasticnet] \\ 
		& alpha  & [$10^{-6}$, $10^{-5}$, $10^{-4}$, $10^{-3}$, $10^{-2}$] \\
		& max\underline{ }iter & [1, 10, 100, 1000, 10000]\\
		& tol & [$10^{-5}$, $10^{-4}$, $10^{-3}$, $10^{-2}$, $10^{-1}$]\\
		\hline
		\multirow{4}*{PassiveAgressiveClassifier} & C & [0.01, 0.1, 1, 10, 100] \\ 
		& max\underline{ }iter & [1, 10, 100, 1000, 10000]\\
		& tol & [$10^{-5}$, $10^{-4}$, $10^{-3}$, $10^{-2}$, $10^{-1}$]\\
		& Loss  & [hinge, squared\underline{ }hinge] \\
		\hline
		\multirow{5}*{RandomForestClassifier} & n\underline{ }estimators & [1, 10, 100, 1000, 10000]\\
		& criterion & [gini, entropy]\\
		& min\underline{ }samples\underline{ }split & [1, 2, 3, 4, 5]\\
		& min\underline{ }samples\underline{ }leaf & [1, 2, 3, 4, 5]\\
		& max\underline{ }features & [none, auto, sqrt, log2]\\
		\hline
		\multirow{4}*{KNeighborsClassifier} & n\underline{ }neighbors & [2, 4, 6, 8, 10]\\
		& algorithm & [auto, ball\underline{ }tree', 'kd\underline{ }tree', 'brute']\\
		& leaf\underline{ }size & [10, 20, 30, 40, 50]\\
		& p & [1, 2, 3, 4, 5]\\
		\hline
		\multirow{5}*{SVC} & C & [0.01, 0.1, 1, 10, 100]\\
		& kernel & [linear, poly, rbf, sigmoid]\\
		& degree & [1, 2, 3, 4, 5]\\
		& tol & [$10^{-5}$, $10^{-4}$, $10^{-3}$, $10^{-2}$, $10^{-1}$]\\
		& max\underline{ }iter & [1, 10, 100, 1000, 10000]\\
		\hline
		\multirow{3}*{LinearDiscriminantAnalysis} & solver & [svd, lsqr]\\
		& tol & [$10^{-5}$, $10^{-4}$, $10^{-3}$, $10^{-2}$, $10^{-1}$]\\
		& shrinkage & [none, auto]\\
		\hline
	\end{tabulary}
\end{table*}

In order to obtain realistic channel estimation data, we used \gls{usrp} B210 \gls{sdr} platforms and an \gls{ofdm} based transmission systems by making use of the GNURadio \cite{gnuradio} \gls{ofdm} transmitter and receiver baseband processing blocks. These provide the functionality for channel estimation based on the Schmidl and Cox method \cite{Schmidl.1997} using the respective preamble symbols. An OFDM setup with an \gls{fft} size of $64$ is considered with $48$ active subcarriers. The cyclic prefix length is $16$ samples, whereas the bandwidth is $10$ MHz and the carrier frequency is $2.484$ GHz. Each data packet consists of $2$ preamble symbols, which are used for channel estimation and frequency offset estimation at the receiver and $N$ data symbols. Here, packets are transmitted with a periodicity of $10$ms, which yields us a vector of $48$ complex symbols for each received \gls{ofdm} packet (only the active subcarriers are considered for channel equalization). We consider a static setup where all participants do not move during transmitting and receiving. The environment is a mixed office/lab area with a lot of objects and metal walls. We record the respective data for several different locations of Bob and Eve respectively, yielding multiple different constellations of \mbox{Bob-Alice} and \mbox{Eve-Alice} pairs as depicted in fig. \ref{environment}. In total, we acquire data from $L_{total}=10$ different position constellations of the users. Further, the \gls{ai} of Eve $P_{\gls{ai}}$ denotes the amount of packets injected by Eve based on the total number of packets $N=100000$ transmitted by Bob and Eve in addition. In case of a transmission of Eve, we assume, as mentioned in section \ref{attacker_model}, that Bobs regular transmission is interfered by Eve (e.g. using a comparatively higher transmit power). Further, we assume that there is always at least one packet transmitted by each of the users within the $N_{train}$ training samples.

\subsection{Processing Pipeline}
In order to finally derive the performance of the classifiers, their hyperparameters need to be selected first. In table \ref{param_space}, an overview of the whole grid of hyperparameter options for all of the investigated classifiers is given. We execute the grid search step as introduced in section \ref{gridsearch} on $L_{valid}=2$ data sets in order to find the optimal hyperparameters $\lambda^*$ for each classifier respectively. These are listed in table \ref{optimal_params}. Finally, the classifiers are parameterized with these sets and we use $L_{test}=8$ testing data sets in order to derive their classification performance for the task of \gls{physec} based message authentication.
\begin{table}[b]
	\caption{Validation and testing parameters}
	\label{general_params}
	\footnotesize
	\centering
	\begin{tabulary}{1.0\textwidth}{|l|l|l|}
		\hline
		Parameter & Variable & Value \\
		\hline
		No. of data sets & $L_{total}$ & 10 \\
		\hline
		No. of validation data sets & $L_{valid}$ & 2 \\
		\hline
		No. of testing data sets & $L_{test}$ & 8 \\
		\hline
		original feature dimension & $M$ & 48 \\
		\hline
		Reduction method & - & mean \\
		\hline
		No. of training samples & $N_{train}$ & 10 \\
		\hline
		Window size & $W$ & 5 \\
		\hline
		Attack intensity & $P_{AI}$ & 25\% \\
		\hline
		No. of samples in each data set & $N$ & 100000 \\
		\hline
	\end{tabulary}
\end{table}

\begin{table}[t!]
	\caption{Optimal hyperparameters of each classifier}
	\label{optimal_params}
	\footnotesize
	\centering
	\begin{tabulary}{1.0\textwidth}{|c|c|c|}
		\hline
		Model & Parameter & Options \\
		\hline
		\multirow{7}*{SGDClassifier} & loss & log \\ 
		& Penalty  & elasticnet \\
		& alpha & $10^{-2}$ \\
		& l1\underline{ }ratio & 1\\
		& max\underline{ }iter & 10000\\
		& tol & $10^{-5}$\\
		& learning\underline{ }rate & adaptive\\
		& eta0 & 0.5\\
		\hline
		\multirow{4}*{PassiveAgressiveClassifier} & C & 0.1 \\ 
		& max\underline{ }iter & 1\\
		& tol & $10^{-5}$\\
		& Loss  & hinge \\
		\hline
		\multirow{5}*{RandomForestClassifier} & n\underline{ }estimators & 100\\
		& criterion & entropy\\
		& min\underline{ }samples\underline{ }split & 3\\
		& min\underline{ }samples\underline{ }leaf & 1\\
		& max\underline{ }features & log2\\
		\hline
		\multirow{4}*{KNeighborsClassifier} & n\underline{ }neighbors & 2\\
		& algorithm & auto\\
		& leaf\underline{ }size & 10\\
		& p & 2\\
		\hline
		\multirow{5}*{SVC} & C & 0.1\\
		& kernel & linear\\
		& degree & 1\\
		& tol & $10^{-5}$\\
		& max\underline{ }iter & 10\\
		\hline
		\multirow{3}*{LinearDiscriminantAnalysis} & solver & [svd, lsqr]\\
		& tol & $10^{-5}$\\
		& shrinkage & auto\\
		\hline
	\end{tabulary}
\end{table}

\captionsetup[subfigure]{labelformat=empty}
\begin{figure*}[t]
	\centering
	\caption{Accuracy metrics for detection of the true transmitter for all classifiers under different side constraints}
	\label{result_figures_all}
	\subfloat[]{\includegraphics[width=\textwidth]{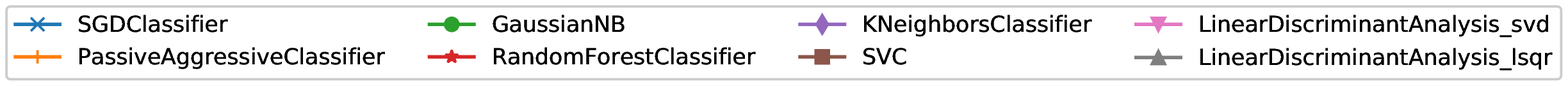}
		\label{legend}
	}
	\vfill
	\subfloat[(a) Detection accuracy under different Attack Intensites]{\includegraphics[width=0.48\textwidth]{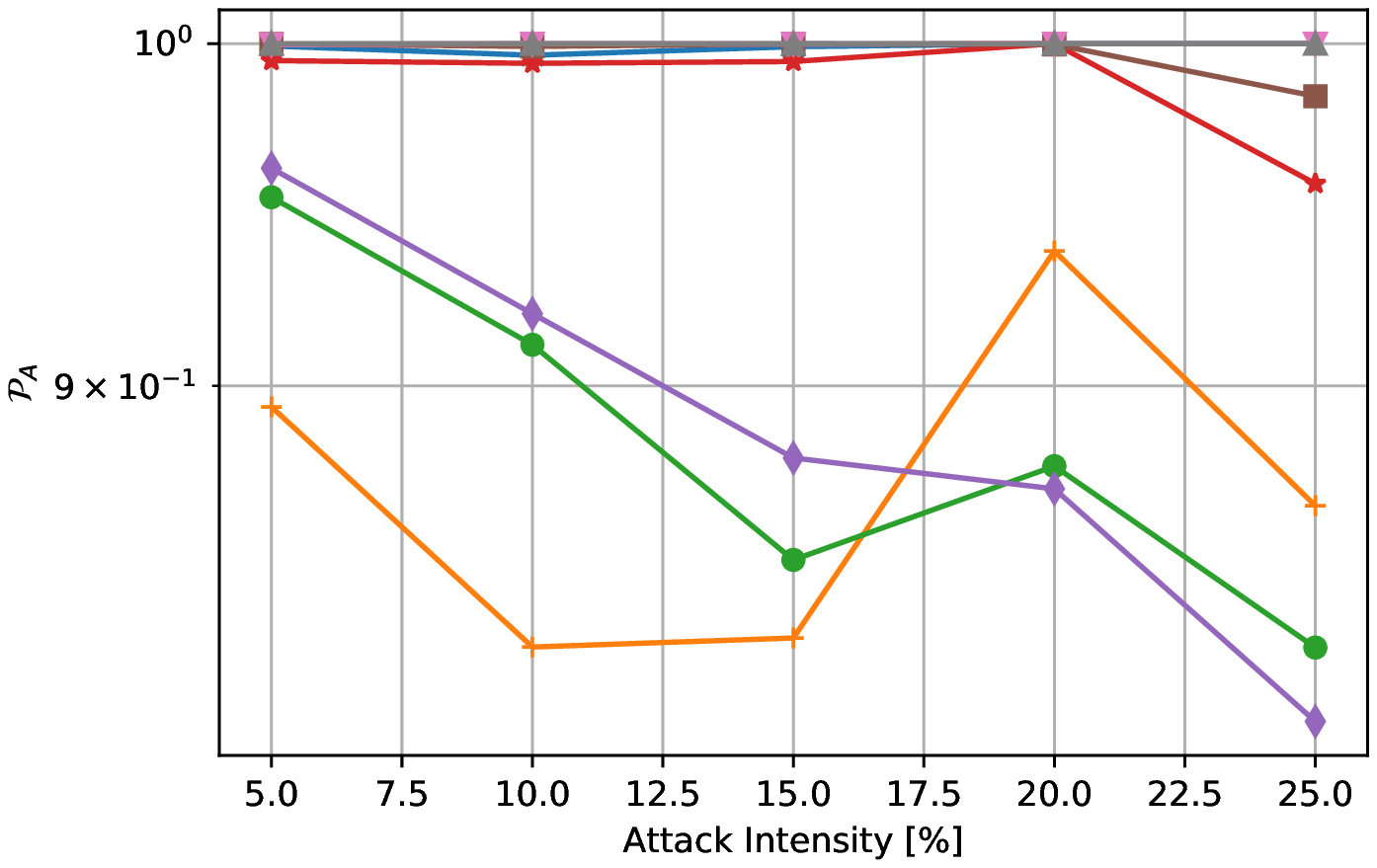}
	\label{ai_accuracy}
	}
	\hfill
	\subfloat[(b) Detection accuracy under different feature dimensions]{\includegraphics[width=0.48\textwidth]{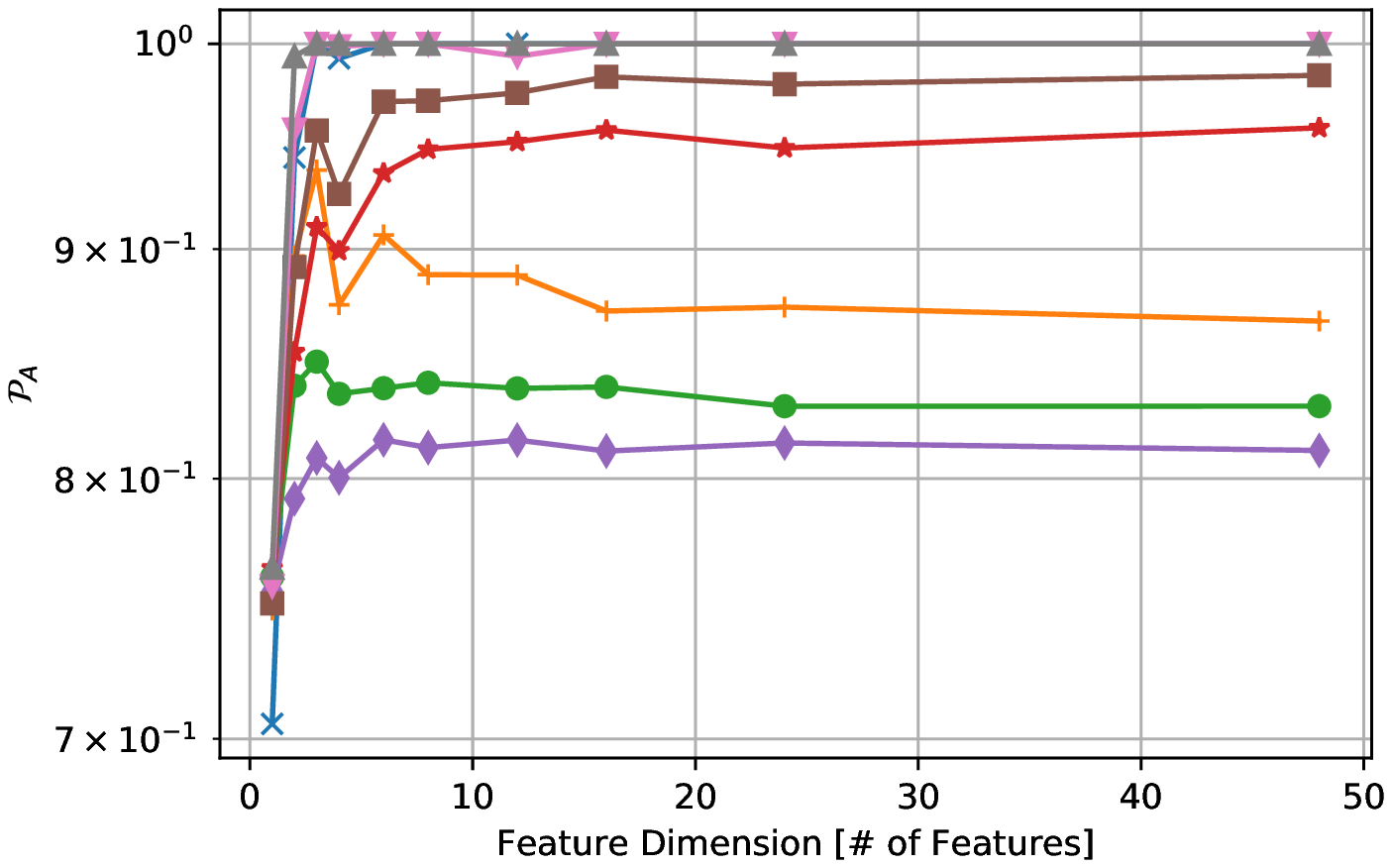}
	\label{feature_dim_accuracy}
	}
	\vfill
	\subfloat[(c) Detection accuracy under different train vector sizes]{\includegraphics[width=0.48\textwidth]{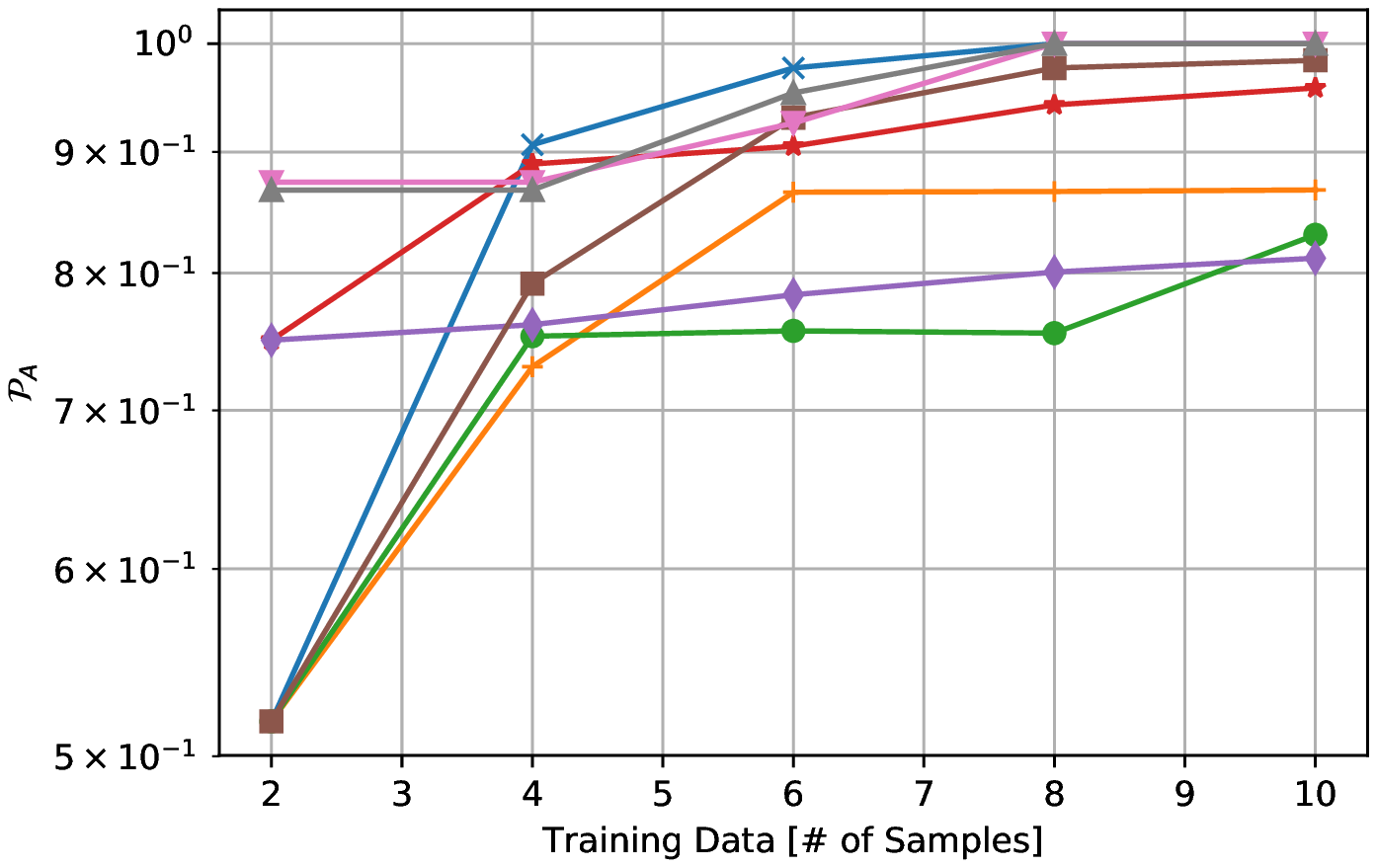}
	\label{train_size_accuracy}
	}
	\hfill
	\subfloat[(d) Detection accuracy under different window sizes]{\includegraphics[width=0.48\textwidth]{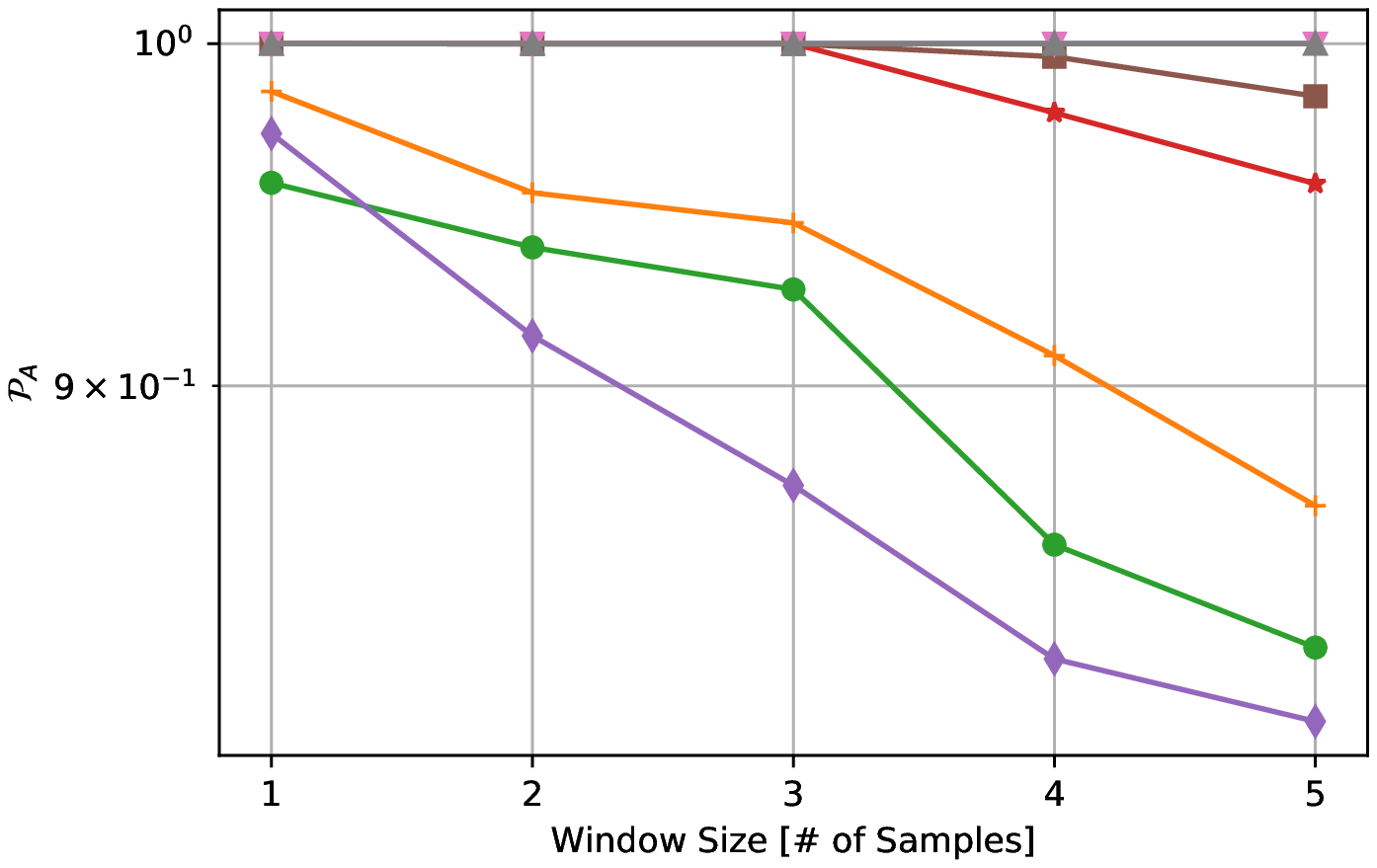}
	\label{window_size_accuracy}
	}
\end{figure*}


\subsection{Results}
Fig. \ref{result_figures_all} shows the performance of the classifier candidates depending on different side conditions. In fig. \ref{ai_accuracy}, the \gls{ai} was changed within a range between $5\%$ and $25\%$. Some classifiers such as the \gls{sgd}, \gls{rf}, \gls{svc} and \gls{lda} perform quite well under varying \gls{ai}s, whereas the performance of the \gls{gnb} and \gls{kn} decrease with increasing \gls{ai}s. Fig. \ref{feature_dim_accuracy} shows the performance of the classifier candidates under different feature dimension using the feature reduction scheme according to (\ref{mean_reduction}). Here, the \gls{sgd} and \gls{lda} perform the best with an accuracy of $100\%$ with at least $16$ features. The performance of all other candidates saturates at specific limits, e.g. the \gls{kn}s performance reaches a maximum of $81\%$ accuracy from a feature dimension of $6$. If a different amount for the training samples is considered, all candidates have a poor performance of below $90\%$ when only $2$ training data samples are used. The performance of each classifier increases, if more training data is used. E. g. the performance of the \gls{sgd} and \gls{lda} both reach $100\%$ accuracy, if $8$ or more training samples are used, whereas the performance of the \gls{svc} does not further increase after $6$ training samples and keeps constant at $86\%$. In fig. \ref{window_size_accuracy}, the dependence of the performance on different window sizes used within \ref{window_function} is shown. Here, the performance of all classifiers does not increase, by taking more past samples into account. The performance of the \gls{sgd} and \gls{lda} stays constant at $100\%$ independent of the window size $W$, whereas the performance of the \gls{rf} and \gls{svc} keeps at that value at least until a window size of $3$ and decreases for higher window sizes $W\ge4$.




\section{Conclusion and Future Work}
\label{CONC}
Under specific scenarios, the application of supervised learning algorithms for channel based message authentication is a promising alternative to conventional security measures such as \gls{mac}. Especially within \gls{urllc} services, a considerable amount of radio resource overhead can be saved. Though the measured performance of some classifiers is comparable to the performance of conventional schemes in our scenario, it also needs to be validated for various other scenarios, such as e. g. mobile setups. Further, other features that are used in addition to channel estimates might yield more robust PHYSEC schemes in order to achieve the goal of message authentication and integrity checking. Another point for future work is to estimate Eves training data, as attackers might not be present yet during the initialization stage. Further, more advanced attacks strategies might be investigated in order to test the robustness of the detection schemes.

\section*{Acknowledgment}
A part of this work has been supported by the Federal Ministry of Education and Research of the Federal Republic of Germany (BMBF) in the framework of the project 16KIS0267 5GNetMobil. The authors alone are responsible for the content of the paper.
 


\bibliographystyle{IEEEtran}
\bibliography{arch_references, PHYSEC_all}

\begin{thebibliography}{10}
\providecommand{\url}[1]{#1}
\csname url@samestyle\endcsname
\providecommand{\newblock}{\relax}
\providecommand{\bibinfo}[2]{#2}
\providecommand{\BIBentrySTDinterwordspacing}{\spaceskip=0pt\relax}
\providecommand{\BIBentryALTinterwordstretchfactor}{4}
\providecommand{\BIBentryALTinterwordspacing}{\spaceskip=\fontdimen2\font plus
\BIBentryALTinterwordstretchfactor\fontdimen3\font minus
  \fontdimen4\font\relax}
\providecommand{\BIBforeignlanguage}[2]{{%
\expandafter\ifx\csname l@#1\endcsname\relax
\typeout{** WARNING: IEEEtran.bst: No hyphenation pattern has been}%
\typeout{** loaded for the language `#1'. Using the pattern for}%
\typeout{** the default language instead.}%
\else
\language=\csname l@#1\endcsname
\fi
#2}}
\providecommand{\BIBdecl}{\relax}
\BIBdecl

\bibitem{AES_standard}
NIST, ``{Advanced Encryption Standard (AES)},'' 2001.

\bibitem{RFC_CMAC.2006}
IETF, \emph{RFC 4493, The AES-CMAC Algorithm}, 2006.

\bibitem{RFC_HMAC.1997}
------, \emph{RFC 2104, HMAC: Keyed-Hashing for Message Authentication}, 1997.

\bibitem{Bockelmann2017}
C.~Bockelmann, A.~Dekorsy, A.~Gnad, L.~Rauchhaupt, A.~Neumann, D.~Block,
  U.~Meier, J.~Rust, S.~Paul, F.~Mackenthun \emph{et~al.}, ``Hiflecs:
  Innovative technologies for low-latency wireless closed-loop industrial
  automation systems,'' \emph{22. VDE-ITG-Fachtagung Mobilkommunikation}, 2017.

\bibitem{Osman.2015}
N.~C.~Y. Osman, Y.-P.~E. Wang, N.~A. Johansson, N.~Brahmi, S.~A. Ashraf, and
  J.~Sachs, ``Analysis of ultra-reliable and low-latency 5g communication for a
  factory automation use case,'' in \emph{2015 IEEE International Conference on
  Communication workshop: London, United Kingdom}, 2015.

\bibitem{Jorswieck.2015}
E.~Jorswieck, S.~Tomasin, and A.~Sezgin, ``Broadcasting into the uncertainty:
  Authentication and confidentiality by physical-layer processing,''
  \emph{Proceedings of the IEEE}, vol. 103, no.~10, pp. 1702--1724, Oct 2015.

\bibitem{Guillaume.}
R.~Guillaume, F.~Winzer, A.~Czylwik, C.~T. Zenger, and C.~Paar, ``Bringing
  phy-based key generation into the field: An evaluation for practical
  scenarios,'' in \emph{IEEE 82nd Vehicular Technology Conference (VTC Fall)},
  2015.

\bibitem{Zenger.2014}
C.~T. Zenger, M.-J. Chur, J.-F. Posielek, C.~Paar, and G.~Wunder, ``A novel key
  generating architecture for wireless low-resource devices,'' in
  \emph{International Workshop on Secure Internet of Things (SIoT)}, 2014.

\bibitem{Ambekar.2012b}
A.~Ambekar, M.~Hassan, and H.~D. Schotten, ``Improving channel reciprocity for
  effective key management systems,'' in \emph{International Symposium on
  Signals, Systems and Electronics (ISSSE), Potsdam, Germany}, 2012.

\bibitem{Xiao.2007}
{L. Xiao, L. Greenstein, N. Mandayam and W. Trappe}, ``Fingerprints in the
  ether: Using the physical layer for wireless authentication,'' in \emph{IEEE
  International Conference on Communications (ICC): Glasgow, Scotland}, 2007.

\bibitem{Xiao.2008}
L.~Xiao, L.~Greenstein, N.~Mandayam, and W.~Trappe, ``Using the physical layer
  for wireless authentication in time-variant channels,'' \emph{IEEE
  Transactions on Wireless Communications}, vol.~7, no.~7, pp. 2571--2579,
  2008.

\bibitem{Pei.2014}
C.~Pei, N.~Zhang, X.~S. Shen, and J.~W. Mark, ``Channel-based physical layer
  authentication,'' in \emph{IEEE Global Communications Conference (GLOBECOM)},
  2014.

\bibitem{Gulati.2013}
N.~Gulati, R.~Greenstadt, K.~R. Dandekar, and J.~M. Walsh, ``Gmm based
  semi-supervised learning for channel-based authentication scheme,'' in
  \emph{IEEE Vehicular Technology Conference (VTC Fall)}, 2013.

\bibitem{Tugnait.2010}
J.~K. Tugnait and H.~Kim, ``A channel-based hypothesis testing approach to
  enhance user authentication in wireless networks,'' in \emph{International
  Conference on COMmunication Systems and NETworks (COMSNETS 2010)}, 2010.

\bibitem{Caparra2016}
G.~Caparra, M.~Centenaro, N.~Laurenti, S.~Tomasin, and L.~Vangelista,
  ``Energy-based anchor node selection for iot physical layer authentication,''
  in \emph{Communications (ICC), 2016 IEEE International Conference on}.\hskip
  1em plus 0.5em minus 0.4em\relax IEEE, 2016, pp. 1--6.

\bibitem{Shi.2013}
L.~Shi, M.~Li, S.~Yu, and J.~Yuan, ``Bana: Body area network authentication
  exploiting channel characteristics,'' \emph{IEEE Journal on Selected Areas in
  Communications}, vol.~31, no.~9, pp. 1803--1816, 2013.

\bibitem{Refaey.2014}
A.~Refaey, W.~Hou, and K.~Loukhaoukha, ``Multilayer authentication for
  communication systems based on physical-layer attributes,'' \emph{Journal of
  Computer and Communications}, vol.~2, no.~8, pp. 64--75, 2014.

\bibitem{Baracca2012}
P.~Baracca, N.~Laurenti, and S.~Tomasin, ``Physical layer authentication over
  mimo fading wiretap channels,'' \emph{IEEE Transactions on Wireless
  Communications}, vol.~11, no.~7, pp. 2564--2573, 2012.

\bibitem{Forssell2019a}
H.~Forssell, R.~Thobaben, and J.~Gross, ``Performance analysis of distributed
  simo physical layer authentication,'' 02 2019.

\bibitem{Forssell2019b}
H.~Forssell, R.~Thobaben, H.~Al-Zubaidy, and J.~Gross, ``Physical layer
  authentication in mission-critical mtc networks: A security and delay
  performance analysis,'' \emph{IEEE Journal on Selected Areas in
  Communications}, 02 2019.

\bibitem{gnuradio}
{GNU Radio}, ``{GNU Radio} website,'' \url{http://gnuradio.org/}, accessed:
  2019-03-22.

\bibitem{Schmidl.1997}
T.~M. Schmidl and D.~C. Cox, ``Robust frequency and timing synchronization for
  ofdm,'' \emph{IEEE Transactions on Communications}, vol.~45, no.~12, pp.
  1613--1621, 1997.

\end{thebibliography}


%

\end{document}